# DiskFiltration: Data Exfiltration from Speakerless Air-Gapped Computers via Covert Hard Drive Noise

(https://www.youtube.com/watch?v=H7lQXmSLiP8)

Mordechai Guri, Yosef Solewicz, Andrey Daidakulov, Yuval Elovici

Ben-Gurion University of the Negev

Cyber Security Research Center

gurim@post.bgu.ac.il; yosef.solewicz@gmail.com; daidakul@post.bgu.ac.il; elovici@post.bgu.ac.il

*Abstract*

Air-gapped computers are disconnected from the Internet physically and logically. This measure is taken in order to prevent the leakage of sensitive data from secured networks. In the past, it has been shown that malware can exfiltrate data from air-gapped computers by transmitting ultrasonic signals via the computer's speakers. However, such acoustic communication relies on the availability of speakers on a computer.

In this paper, we present 'DiskFiltration,' a covert channel which facilitates the leakage of data from an air-gapped compute via acoustic signals emitted from its hard disk drive (HDD). Our method is unique in that, unlike other acoustic covert channels, it doesn't require the presence of speakers or audio hardware in the air-gapped computer. A malware installed on a compromised machine can generate acoustic emissions at specific audio frequencies by controlling the movements of the HDD's *actuator* arm*.* Digital Information can be modulated over the acoustic signals and then be picked up by a nearby receiver (e.g., smartphone, smartwatch, laptop, etc.). We examine the HDD anatomy and analyze its acoustical characteristics. We also present signal generation and detection, and data modulation and demodulation algorithms. Based on our proposed method, we developed a transmitter on a personal computer and a receiver on a smartphone, and we provide the design and implementation details. We also evaluate our covert channel on various types of internal and external HDDs in different computer chassis and at various distances. With DiskFiltration we were able to covertly transmit data (e.g., passwords, encryption keys, and keylogging data) between air-gapped computers to a smartphone at an effective bit rate of 180 bits/minute (10,800 bits/hour) and a distance of up to two meters (six feet).

## 1. Introduction

An 'air-gap' is a measure taken in order to keep a computer network (or other type of IT devices) disconnected from public networks such as the Internet. In air-gap isolation, there is no wired or wireless connection between the internal network and the outer world. Given the high level of separation, attackers cannot breach the network and steal data using remote attacks launched over the Internet. Military networks, as well as networks within financial organizations, critical infrastructure, and commercial industries [1] [2], are known to be air-gapped due to the sensitive data they store and process. Despite the high level of isolation, an air-gap doesn't provide hermetic protection from breach events. Several incidents in which air-gapped networks have been compromised have been published in the recent years [3] [4] [5] [6]. Infecting such networks can be accomplished through a malicious insider, stolen credentials, physical access, and so on [7].

Once the attacker has a foothold in the target network, he/she may want to exfiltrate valuable data. To that end, the attacker has to overcome the physical isolation by *bridging the air-gap*. Over the years, different types of covert channels have been proposed by security researchers, enabling exfiltration through an air-gap. Electromagnetic methods that exploit electromagnetic radiation from different components of the computer [8] [9] [10] [11] are likely the oldest type of covert channel researched. Various types of optical [12], thermal [13], and acoustic [14] [15] [16] out-of-band communication channels have also been suggested.

## 1.1 Speakerless Computers

Most of the acoustic covert channels require a speaker (as a transmitter) and a microphone (as a receiver) to be installed in the air-gapped computers, in order to enable bi-directional covert communication. A malware can encode the data over sonic or ultrasonic frequencies, and subsequently broadcast it through the computer speaker. Another computer with a microphone can receive the transmissions, decode the data, and sent it to the attacker. To avoid such an attack, security policies may prohibit the use of speakers and microphones in a secure network, a measure also referred as an 'audio-gap' [17] [18]. Keeping speakers disconnected from sensitive computers can effectively mitigate the acoustic covert channels based on speakers [19].

In this paper, we introduce 'DiskFiltration,' an acoustic channel which works even when speakers (or other audio related hardware) are not present in the infected computer. Our method is based on exploring intrinsic covert noises emitted from the hard disk drive (HDD) which exists on most computers today. We show that malicious code on a compromised computer can perform 'seek' operations, such as the HDDs moving head (the actuator) will induce the generation of such noise patterns at a certain frequency range. Arbitrary binary data can therefore be modulated through these acoustic signals, and the signals can then be received by a nearby device equipped with a microphone (e.g., smartphone, smartwatch, or laptop), and be decoded and finally sent to the attacker.

## 1.2 HDD, SSD, and SSHD

Three primary types of mass storage drives exist today:

1. **Hard Disk Drive (HDD).** The hard disk drive uses a mechanical arm (actuator) with a read and write head to access information at the correct location on a spinning magnetic platter**.** The hard disk drive is the most prevalent mass storage medium used in PCs, servers, legacy systems, and laptops [20].

2. **Solid State Drive (SSD).** Solid state drives store the data in interconnected flash memory chips (e.g., NAND based flash chips), a type of non-volatile memory. There are no moving or mechanical parts to an SSD, and hence they emit virtually no noise.

3. **Solid State Hard Drive (SSHD).** Solid state hybrid drives (SSHDs) combine HDD and SSD technology in the same unit. The flash is used as a cache buffer for frequently used data, while the rest of the data is stored on the magnetic media. SSHD contains the HDD's mechanical parts, and hence emits noise when the data stored on the HDD component is accessed.

Our method is based on the acoustic signals generated by the hard drive's mechanical parts, and therefore is relevant to HDDs and SSHDs, as opposed to SSDs. Generally speaking, SSDs are considered to have advantages over HDDs in term of data access speed, unit size, and reliability. Despite the increased rate of adoption of SSDs, HDD are still the most sold storage devices, mainly due to their low cost. In 2015, 416 million HDD units were sold worldwide, compared to 154 million SSD units. Currently, HDDs still dominate the storage wars, and most PCs, servers, legacy systems, and laptops are installed with HDD drives [20]. This means that our covert channel is available on most of today's desktop and servers.

The rest of this paper is structured as follows. In section 2 we present related work. Section 3 discusses the attack model. Section 4 introduces the anatomy of hard disk drives, and section 5 discusses its acoustic characteristics. Section 5 describes the implementation of the transmitter and receiver. Section 6 present evaluation results. Section 7 proposes countermeasures, and we conclude in section 8.

## 2. Related Work

Covert channels allowing exfiltration of data from air-gapped computers can be categorized into electromagnetic, optical, thermal, and acoustic. The general technique of spying on information systems through leaking emanations is also referred as a TEMPEST attack [21] [22].

- **Electromagnetic.** Electromagnetic emanations from different computer components have been investigated as a medium for data transmission for more than twenty years. Intentional emissions from a computer screen was first discussed by Kuhn and Anderson [8] and Thiele [23]. More recently, AirHopper malware [24] used the video cable to generate FM radio transmissions, in order to leak data to a nearby mobile phone. In the same manner, GSMem [11] and Funthenna [25] exploit electromagnetic radiation generated from the computer bus to transmit data over the air-gap. Kasmi et al. proposed injecting command and control by using electromagnetic interferences [26] [27]. Other types of electromagnetic based methods are discussed in [28] [29] [30].

- **Optic.** Optical methods are less discussed in the context of covert channels, since they are visible to the surrounding environment. Data leakage through keyboard LEDs was proposed in [12] and through office scanner lights in [31]. VisiSploit, a covert optical method, was proposed by Guri et al [32]; in this method, data is leaked from the LCD screen to a remote camera via an invisible image projected on the screen.

- **Thermal**. Air-gap communication using heat emissions was proposed in [13]. In a method called BitWhisper, the authors demonstrated slow communication between adjacent air-gapped computers via heat exchange. Thermal covert channels on modern multicores (in the same system) have been thoroughly studied by Bartolini, Miedl and Thiele [33].

- **Acoustic.** Acoustic methods are based on leaking data over sound waves at sonic and ultrasonic frequencies. Data transmission over audio was first reviewed by

Madhavapeddy et al. in 2005 [34] when they discussed audio based communication between two computers. In 2013, Hanspach and Goetz [35] used near-ultrasonic soundwaves to establish a covert channel between air-gapped systems equipped with speakers and microphones. They implemented a botnet which communicated between computers at distance of 19.7 meters with a bandwidth of 20 bits/second. O'Malley and Choo have examined different exfiltration scenarios using laptop speakers and microphones at high frequency sounds up to approximately 23kHz [36]. The concept of communicating over inaudible sounds has been comprehensively examined by Lee et al [15]. The work in [37] extends the ultrasonic covert channel for smartphones, demonstrating how data can be transferred up to 30 meters away. Interestingly, in 2013, security researchers claimed to find BIOS level malware in the wild (dubbed BadBios) which communicates between air-gapped laptops using ultrasonic sound [38].

Notably, the aforementioned acoustic methods assume installation of audio hardware (including speakers) within the transmitting PC or laptop. However, speakers are sometimes forbidden from certain computers based on regulations and security practices [17]. In 2013, Genkin et al. described a so-called acoustic cryptanalysis key extraction attack, which enables the reconstruction of encryption keys from noise generated by the CPU [39]. In 2016, Guri et al introduced Fansmitter, a malware which facilitates the exfiltration of data from an air-gapped computer via noise intentionally emitted from the CPU and chassis fans [40]. In this method, the computer does not need to be equipped with audio hardware or an internal or external speaker. Methodologies, mitigation, and measurements for out-of-band covert channels are provided in [41] [42] [43]. Table 1 summarizes the different types of covert channels for air-gapped computers, including the acoustic channels.

Table 1. Different types of air-gap covert channels

| Method | Examples |
|---|---|
| **Electromagnetic** | [8] [9], AirHopper [24], GSMem [11], Funthenna [25] [44] [28] |
| **Optical** | [12] [31], VisiSploit [32] |
| **Thermal** | BitWhisper [13] |
| **Acoustic (via speakers)** | [14] [15] [36] [37] |
| **Acoustic (without speakers)** | Fansmitter [40], DiskFiltration (this paper) |

# 3. Attack Model

DiskFiltration, as an acoustic covert channel, can be used to leak data from air-gapped computers. However, this covert channel can also be used in the case of Internet connected computers (non air-gapped) in which the network traffic is intensively monitored by networked-based intrusion detection (IDS), intrusion prevention (IPS) and data leakage prevention (DLP) systems. In these cases, exfiltration of data though the Internet traffic may be detected, and hence the attacker may want to resort to an out-of-band covert channel.

The adversarial attack model consists of a *transmitter* and a *receiver*. The transmitter is usually an ordinary desktop computer or server with at least one HDD installed. The receiver is a nearby device with audio recording capabilities. It can be a smartphone placed on the table, a smartwatch on the user's hand, or a nearby laptop with a microphone (Figure 1). Infecting highly secure networks can be accomplished, as demonstrated by incidents such as Stuxnet [45], Agent.Btz [46], and others [47] [48] [7]. Infecting a mobile phone or other recording device can be accomplished via different attack vectors, using emails, SMS/MMS, malicious apps, and so on [49] [50] [51] [52].

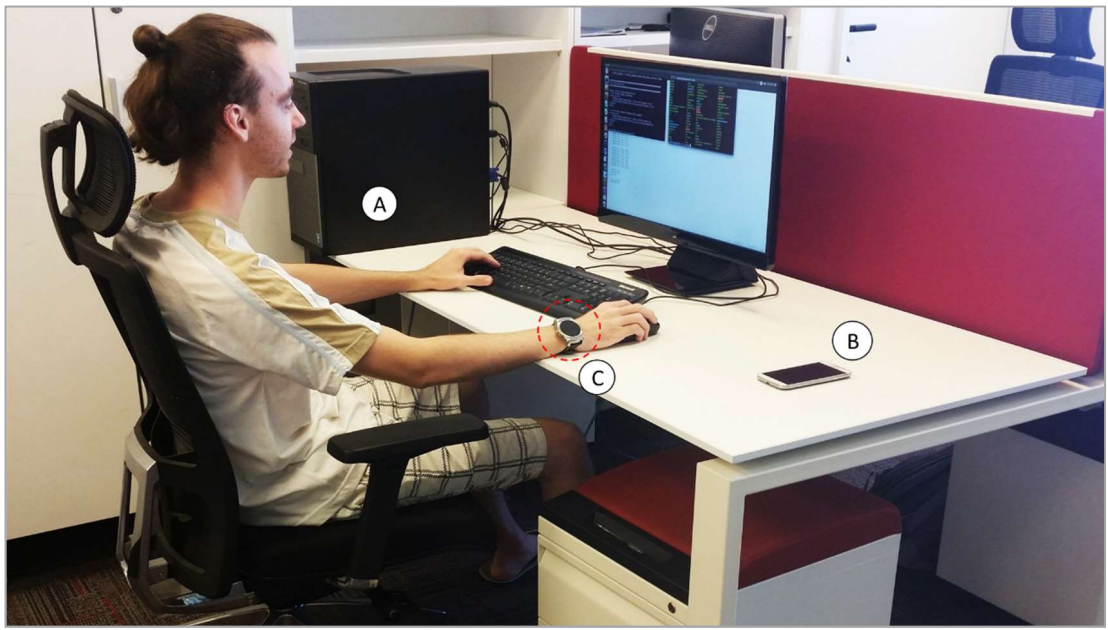

**Figure 1. A typical exfiltration scenario.** A compromised computer (A) - without speakers, and with audio hardware disabled - transmits sensitive information via covert acoustic signals generated by its hard disk drive. This information is received and decoded by a nearby mobile phone (B), smartwatch (C), laptop, or other device with recording capabilities.

The malware installed on the computer gathers the data to exfiltrate (e.g., passwords or encryption keys), and then transmits it using acoustic signals emitted from the HDD. The acoustic signals are generated by performing intentional seek operations which cause the HDD actuator arm to make mechanical movements. The nearby receiver receives the transmission, decodes the data, and transfers it to the attacker via mobile data, SMS, or Wi-Fi.

## 4. Anatomy of a Hard Disk Drive

In this section we provide the technical background necessary to understand the way DiskFiltration works. A more comprehensive description of HDD functionality and its internal operation can be found in [53].

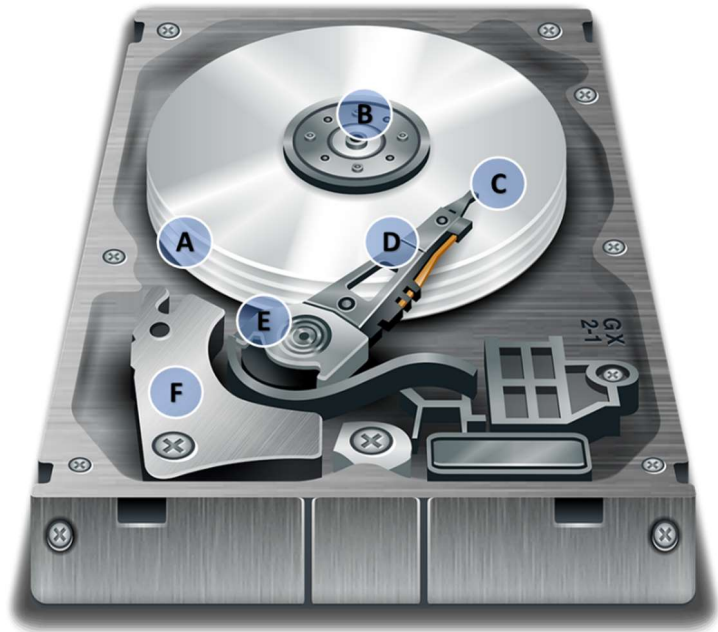

**Figure 2. A hard disk drive's internal parts**

The internal view of a hard disk drive is shown in Figure 2. Hard disk drives store data in disks, or *platters*, coated with magnetic material (Figure 2, A). The platters rotate at various speeds, depending on the type of HDD. Modern consumer-grade HDDs commonly have rotational speeds of 5400, 7200, or 15,000 revolutions per minute (RPM). The engine that rotates the platters is the *spindle motor* (Figure 2, B), and it spins at a constant speed that is tied to the RPM of the HDD. Notably, this motor is one of the constant sources of noise from a HDD. The magnetic data is read/written from/to the platters using *read-and-write heads* (Figure 2, C). These heads are positioned very close to the magnetic surface (at a distance of nanometers from one another) and can detect (read) or change (write) the magnetization of the material passing under it. Modern HDDs have several stacked platters, each of which has its own read-and-write head. All of the read-and-write heads are attached to the *actuator arm* (Figure 2, D). During read and write operations, the *actuator* (Figure 2, F) rotates the *actuator axis* (Figure 2, E) which moves the read-and-write heads on an arc across the platters as they spin. The mechanical movements of the actuator generate noise at different levels and frequencies. Video clips showing HDD internal parts during operation can be found online [54] [55].

## 4.1 Disk Geometry

The basic geometry of the HDD platter is depicted in Figure 3.

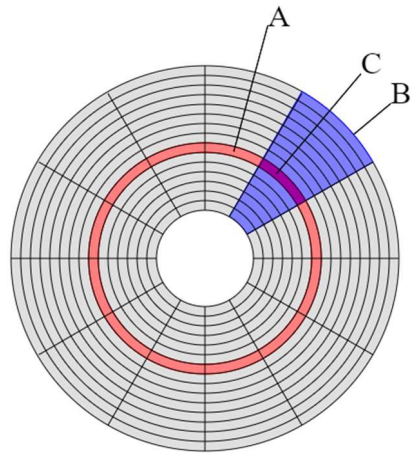

**Figure 3. Basic geometry of an HDD platter**

Modern platters are typically made using an aluminum or glass and ceramic substrate which stores the magnetic data. Each platter has its own read and write head which reads and writes the data from the surface. Physically, the magnetic data is stored on circles on the surface known as *tracks* (Figure 3, A). Corresponding tracks on all surfaces of a drive (on all platters) make up a *cylinder*. Two fundamental terms of disk geometry are the geometrical sector and the disk sector. A *geometrical sector* (Figure 3, B) is a section of a disk between a center, two radii, and a corresponding arc. A *disk sector* (Figure 3, C) refers to the intersection of a track and geometrical sector. Logically, the disk sector is the minimum storage unit of a hard drive. A detailed description of disk geometry is provided in [53] [56].

'Seek' describes the operation of the actuator arm to move to a specific track of the disk where the data needs to be read or written. The time it takes to move the head to the desired track is called the *seek time*. As we describe in the following section, the movement of the head assembly on the actuator arm during the seek operation emits acoustic noise.

Today, there are different types of HDDs with various sizes and capacities. The 2.5 inch and 3.5 inch hard disks are the most popular sizes today. The capacity of modern HDDs is between hundreds of gigabytes to few terabytes.

## 5. HDD Acoustics

An HDD emits noise at different frequencies and intensity levels which are produced by the movements of its internal parts. Notably, although there have been several studies on the acoustic characteristics of a hard drive, the noise emission mechanisms and the precise source of such emissions have not been comprehensively modeled [57] [53].

There are two primary sources of acoustic noise inside a drive: the motor and the actuator. These sources correspond with two type of noises as explained below.

**Idle acoustic noise** is defined as the noise generated when the HDD spins the disks (platters). Idle noise is generated mainly by the spindle motor and the ball bearings inside the motor. This main frequency of idle noise can be calculated by $IdleMainFreq = RP\ /\ 60$ where

$RPM$ is the HDD rotation speed. Figure 4 shows the spectrogram of the idle acoustic noise generated by the Western Digital HDD spinning at 7200 RPM. The primary tone is generated at $7200/60 = 120\ Hz$, and can be seen in the spectrogram as a highlighted continuous frequency peak.

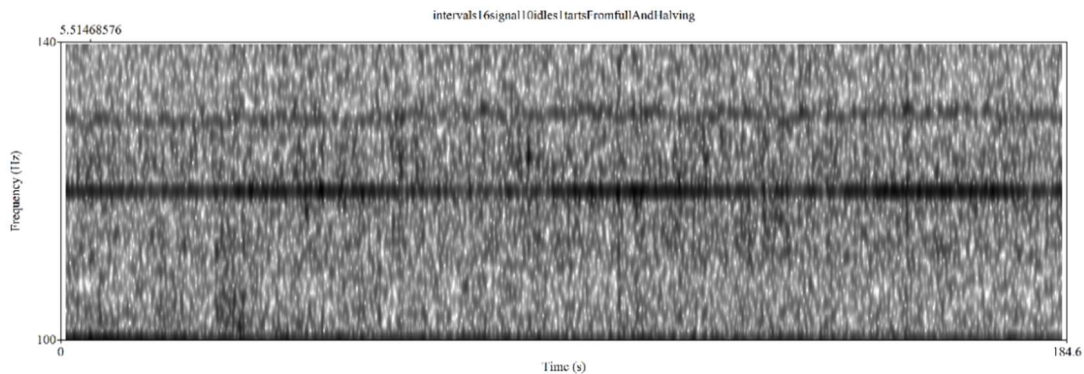

Figure 4. Spectrogram of idle acoustic noise generated by a HDD with an RPM of 7200

**Seek acoustic noise** is generated by the engine of the actuator and its movement during seek, read, and write operations. This noise is produced during file system activities (e.g., file read and write) and is usually louder than the static Idle acoustic noise. Unlike idle acoustic noise, the seek noise depends on many factors (magnetoelectric interactions, vibrations, and so on), hence the exact tone frequency cannot be calculated by a formula [57] [53]. The exact seek tone frequency regions (expected to be up to a range of $6Khz$ [57]) can be investigated through manual or automatic waveform analysis. In this research, we exploit frequency regions which are probably rooted on the hard disk seek time, and in particular, on the shortest seek time component, the track-to-track seek time, which is the time required to move from adjacent tracks. As is shown later, the most informative frequency region detected in our experiments is around 2080 Hz, which is equivalent to a 0.48 ms track-to-track seek time.

## 5.1 Noise Reduction Technologies

Many HDD manufacturers include a feature called automatic acoustic management (AAM) [58] which aims at reducing seek acoustic noise. Such technologies (e.g., Western Digital IntelliSeek [59]) use sophisticated algorithms to regulate the acceleration and positioning of the HDD actuator so that the emitted noise is reduced. Enabling and disabling this feature is possible with the appropriate software or with an API to the HDD firmware [60] [61] [62]. During our experiments we didn't modify the AAM setting, which is usually set to on by default. The main reason we do so is to keep our covert channel as stealth and quiet as possible in order to evade detection by the user.

## 5.2 Acoustic Signal Generation

As explained, the *idle acoustic noise* emitted from disk rotation is static and cannot be controlled by software. In order to modulate binary data, we exploit the *seek acoustic noise* generated by the movements of the actuator. By regulating (starting and stopping) a sequence of seek operations, we control the acoustic signal emitted from the HDD, which in turn can be used to modulate binary '0' and '1.' Next, we examine the *seek acoustic noise* generated by three types of operations: read, write, and seek.

### 5.2.1    'Read' and 'Write' Operations

Figure 5 shows the spectrograms of acoustic waveforms generated from the HDD during read (left image) and write (right image) operations as recorded from outside the computer chassis. In this test we read the content of 100 MB binary file to a buffer in the memory, and write 100MB of random bytes to a file in the disk. During the tests, the cacheing on all levels was bypassed to guarantee physical disk access. Read and write operations cause acoustic bursts (as seen as a general increase in frequency for a short time period), during most of these operations, the head stays at the same position.

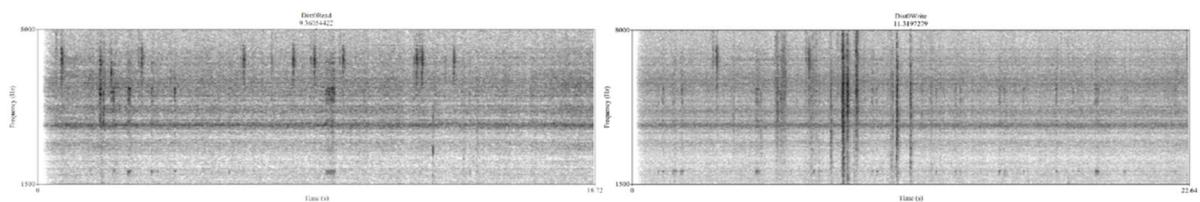

**Figure 5. Spectral views of read (left) and write (right) operations**

### 5.2.2    'Seek' Operations

Figure 6 shows the spectrogram of an acoustical waveform generated from the HDD during seek operations as recorded from outside the computer chassis. In this test we repeatedly cause the head to move between two consecutive tracks by performing seek operations within a loop for a period of three seconds. As can be seen by the highlighted discrete frequency peaks in the spectrogram, during the seek operations there is a strong signal wrap around a range of discrete frequencies between 1500 – 8000 Hz.

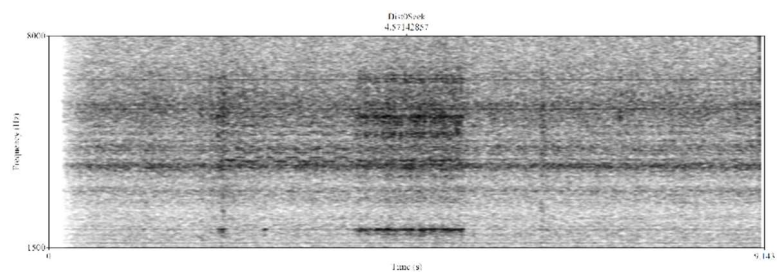

**Figure 6. Spectral view of seek operations**

We also examine the acoustic noise emitted by seek operations when the actuator moves between tracks at different distances. Figure 7 shows the acoustical waveform generated from the HDD during seek operations as recorded from outside the computer chassis. In this test we perform three types of seek operations, (1) seeking and reading repeatedly from the first and last sectors, (2) seeking and reading between two consecutive tracks, and (3) seeking and reading between two consecutive sectors. As can be seen, the seek and read operations cause an acoustic signal to wrap all over the range of 0 to 6000 Hz. There were no significant acoustic differences (frequencies or amplitude) between the three types of seek operations. This indicated that in order to emit a noticeable level of noise it is sufficient to perform seek operations between any two tracks. In our tests we used the seek operation between the first and the last track of the HDD.

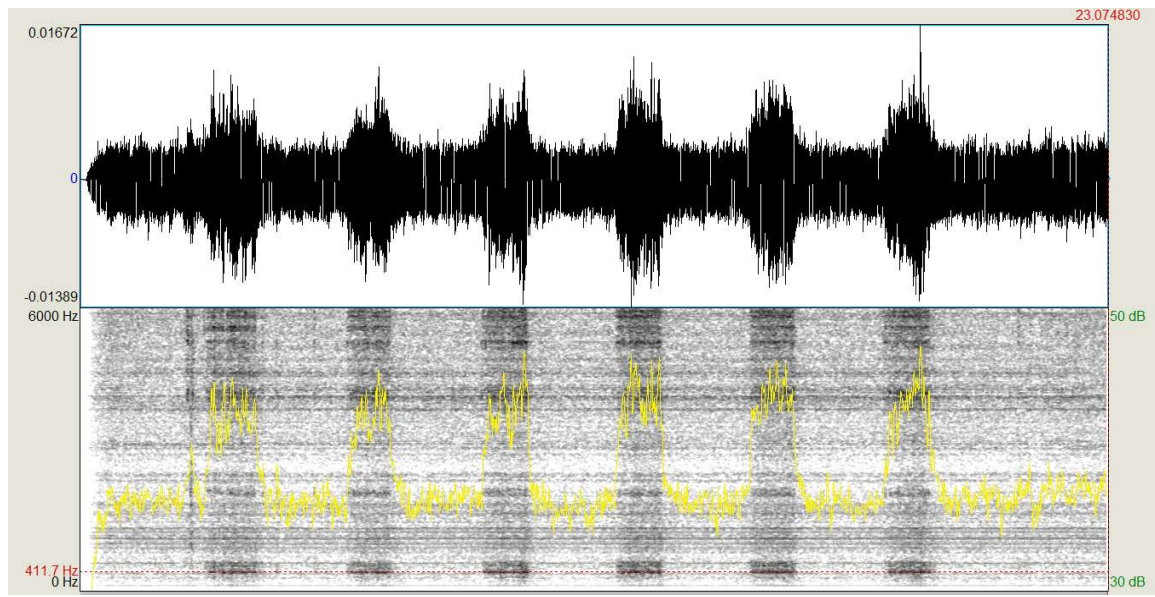

**Figure 7. Spectral view of 'seek' operations between different tracks**

## 5.3 Cache Avoidance

In order to efficiently modulate data over the acoustic signal, we need to control the precise timing and duration of the HDD operations. In particular, we need to be able to instruct the HDD controller to perform an operation at a given time without delay or caching. Modern operating systems, including Linux and Windows, employ disk and files I/O caching mechanisms in their kernel or device drivers. Such caching can cause timing delays and inconsistencies in the generation of the acoustic signals. For efficient and error-free signal generation, the OS must be instructed to avoid any type of caching during the I/O operation. File system cache avoidance is possible in Linux and Windows [63] [64]. In addition to the OS cache, the HDD controllers also use a type of cache. For example, with *write-back cache* enabled, the HDD stores the data to be written in an intermediate buffer rather than in the disk. For precise timing of the signal, controller level cache can be turned off using appropriate tools [61].

# 6. Implementation

In this section we describe the implementation of the DiskFiltation transmitting software, including signal generation, data modulation, and bit-framing. We also describe the implementation of a receiver as an Android app for the smartphone.

## 6.1 Transmitter

A program can perform disk operations with two types of addressing: file system addressing and direct disk addressing. In file system addressing, the running process specifies the file name to perform the read or write operations on. In direct addressing, the process specifies the physical location on the HDD layout for the required I/O operation, e.g., specifying a sector number to read from or write to. Modern OSs, such as Windows and Linux, provide APIs for the two type of addressing; in particular, they allowing direct disk addressing [65] [66]. Technically, it means that user-level processes can generate the acoustic signals by performing seek operations by specifying sector numbers or the location within files. Notably, file level operations may not require any special permissions (e.g., root). For example, any process may be able to read and write files from or to temporary or working folders.

We implemented a prototype of the transmitter for the Linux OS. We choose to use the *seek* operation, as it generates the highest level of the acoustic signal. We implemented a C program which uses the direct addressing system calls using the *fopen(),* and *fseek()* systems' calls [67] [68]. We also used a shell script version of the transmitter using the Linux dd command-line utility [69]. This is a low level utility of Linux which can perform a wide range of HDD operations (e.g., read or copy) at the file or block level.

**Caching:** During the transmissions, we turned off the OS disk cache using the */proc/sys/vm/drop_caches* in order to instruct the kernel to free the pagecache, dentries, and inodes. We also turned off the HDD *write-back cache* mechanism, using the hdparam command line tool [61]. In our shell script, we used the dd with *direct* the flag (using direct I/O for data), and *sync* flag (using synchronized I/O for data).

### 6.1.1 Data Modulation

To transmit binary data we used a simple on-off keying (OOK) modulation. In this digital modulation scheme, data is represented by the presence of a carrier at a specified frequency $Fc$. More specifically, a binary '0' is represented by the presence of a carrier for a duration of $T_1$, while its absence for the duration of $T_0$ represents a binary '0.' Algorithm 1 shows a pseudo code for our C program which handles the transmission of a bit $b$.

```
Algorithm 1   TransmitBit
1: procedure transmitBit(b, T0, T1, BEGIN_SEC, END_SEC)
2:  sync(); //drop cache
3:  hddDev = open(/dev/sda)
4:  if (b='0') then
5:      Sleep (T0);
6:      return;
7:  if (b='1') then
8:      for time T1 do
9:          seek(hddDev, BEGIN_SEC);
10:         seek(hddDev, END_SEC);
11:     end for
12: return;
```

The transmitBit procedure receives the '0' and '1' transmission time (*T0, T1)* and two sector numbers for the seek operation *(BEGIN_SEC, END_SEC*). As we previously explained, in signal generation, moving the actuator between the sectors positioned in different tracks produces the highest level of noise. If the bit to transmit is '0,' the procedure does nothing by sleeping for duration *T0*. If the bit to transmit is '1,' the procedure invokes seek operations, causing the head to repeatedly move between BEGIN_SEC and END_SEC for duration *T1*.

A skeleton of a shell script using dd for the transmission of '1' using the seek operation follows.

```
#!/bin/sh

i=0
a= BEGIN_SEC
b= END_SEC

while [ $i -lt $1 ]
do
   sync
   echo 3 > /proc/sys/vm/drop_caches
   dd if=/dev/sda of=/dev/null skip=$a count=1 bs=512
   dd if=/dev/sda of=/dev/null skip=$b count=1 bs=512
   i=`expr $i + 1`
   a=`expr $a + 10000`
   b=`expr $b + 10000`
done
```

The sample script runs the actuator between two sectors using a simple read operation. Note that the sector numbers are incremented at every iteration. This is done to evade a potential caching mechanism (within the OS or HDD controller). The BEGIN_SEC and BEGIN_SEC have to be selected, because they are positioned on different tracks.

6.1.2   Bit Framing

As explained previously, unlike the idle acoustic noise, seek acoustic noise may vary depending on the type of HDD, and differences in seek acoustic noise can also vary between HDDs of the same model. Although the general range of seek tone frequency is known (e.g., $0 - 6Khz$ [57]), the exact tone frequency cannot be calculated by a formula. This implies that a potential receiver (e.g., an application in a smartphone) needs to scan the frequency range first, in order to find and detect the carrier used for the on-off keying modulation. In addition, $T_0$ and $T_1$ may be set differently on each transmitter (e.g., given the expected SNR), and may be unknown to the receiver in advance. To assist the receiver in dynamically synchronizing with the transmitter parameters, we transmit data in small frames. Each frame consists of a preamble sequence of four bits and a payload of 36 bits (Table 2).

Table 2. A frame consisting of four bits of preamble, followed by a payload of 36 bits

| Preamble (4 bits) | Payload (36 bits) |
|---|---|
| 1010 | 0101110101010101… |

The preamble consists of the '1010' sequence and is used by the receiver to periodically determine the carrier frequency. In addition, the preamble header allows the receiver to identify the beginning of a transmission in the area and extract other channel parameters, such as $T_0$ and $T_1$.

6.1.3   Stealth

As noted, modern HDDs include a feature called AAM [58] which reduces seek acoustic noise. In order to keep the covert channel as stealth as possible, we did not modify the AAM setting, resulting in quiet HDD operation. Our experiments show that in modern HDDs, the generated acoustic signals blend with the background noise and are not noticeable by the user. Users may notice the HDD activity by seeing the HDD's blinking LED or hearing unusual seek noise.

However, such occurrences won't raise suspicions, since they aren't out of ordinary because the HDD is routinely active due to swapping, indexing, backups, and other types of background operations.

## 6.2 The Receiver

Directly decoding the acoustic information from the transmitted waveform is not efficient, since the relevant information encoded by the induced HDD operations are concentrated in narrowband frequency regions. The signal-to-noise ratio (SNR) of the captured waveform can be significantly improved by exploiting the informative spectral regions, rather than the whole frequency spectrum. In this research, these regions are defined experimentally, because a theoretical determination of the position of the spectral peaks is quite complex, as discussed earlier.

In order to analyze our distinct encoding, we estimated the SNR in the frequency domain (as opposed to the time domain) as follows. Our "signal" (**X**) level is estimated by summing the magnitudes of the FFT bins within our defined informative regions (R) during induced seek operations (bit 1). Noise level (**N**) is estimated in the same way, during an idle noise interval (bit 0). The SNR (in dB) is therefore the logarithm ratio of these two quantities:

$$SNR_R = 20 * log\left(\sum_{k=R} |X_k| \Big/ \sum_{k=R} |N_k|\right)$$

The signal adds up coherently in the frequency domain, whereas noise adds up incoherently. Therefore, in order to maximize the SNR, we would like to define R encompassing the most informative frequency bins. Note that the windowing settings and number of spectral bins used should be optimized in order to avoid spectral leaking (single frequencies spread through adjacent bins) and improve the SNR characteristics.

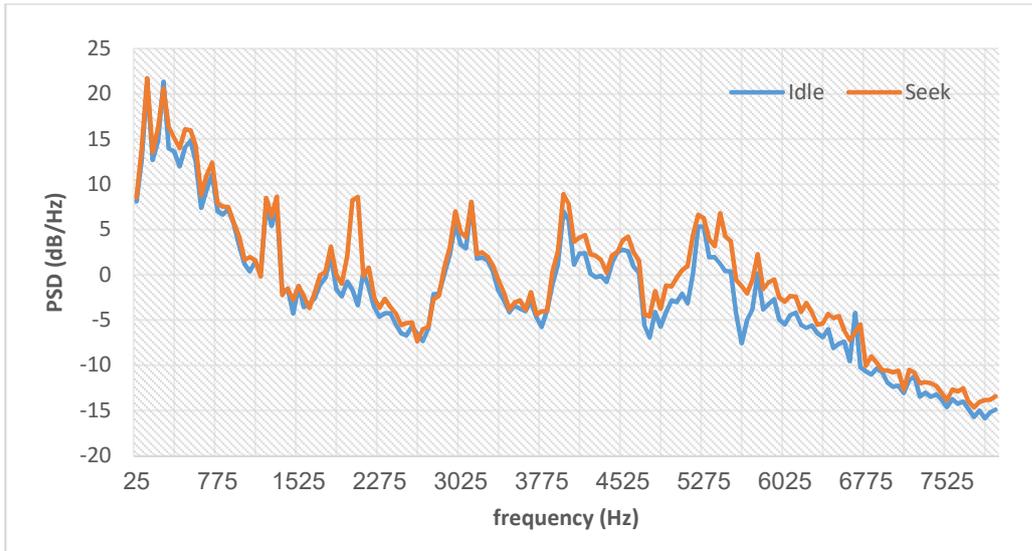

Figure 8. PSD density for 'seek' and 'idle' signals

Figure 8 depicts the power spectral density (PSD) for seek and idle wave excerpts from the seek and read operations. The PSD reflects the average power of the signal in a logarithmic scale during a specific time-frequency region. It is expressed in dB relative to the auditory threshold. This wave was captured at a very close distance to the source, at 44.1 kHz, and resampled to 16 kHz. Fast Fourier transform (FFT) was calculated for 160 bins, spanning 50 Hz each. Spectral peaks are clearly spotted in the graphs, and the strongest low frequency peaks correspond to the basic frequency of 120 Hz generated by a 7200 RPM disk. Figure 9 depicts the seek-idle PSD ratio for each frequency bin in greater detail. It can be observed that the 2050-2100 Hz region is the most informative in terms of the SNR. This means that optimal SNR estimation should be focused on this region. For instance, direct SNR calculation on the whole waveform (using all frequency bins) yields an SNR of 1.5 dB, as opposed to 12.0 dB obtained by setting R to 2050-2100 Hz (using the bin corresponding to the highest SNR).

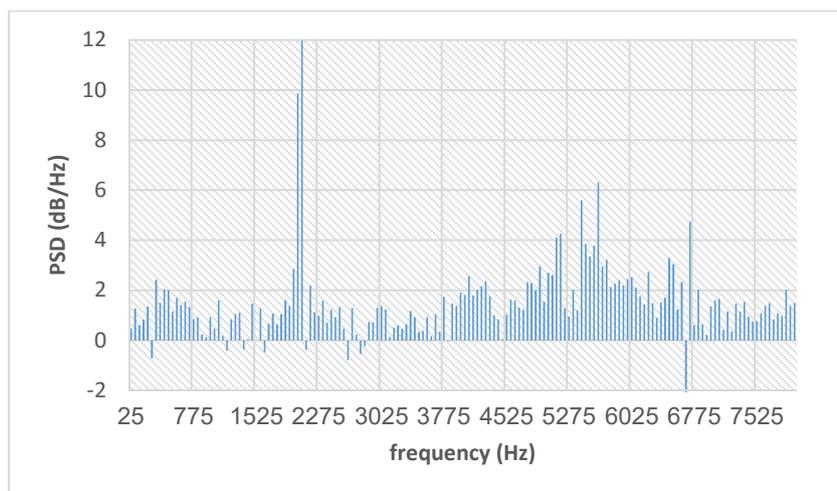

Figure 9. PSD SNR as function of the frequency

### 6.2.1 Signal Decoding

From a signal processing perspective, our decoder can be implemented as an envelope detector of the waveform energy in the above mentioned frequency regions. In particular, we band-pass filter the received waveform between 2050-2100 Hz and then smooth the narrowband signal, convolving with an analysis window in order to estimate its intensity. The window length should be adjusted according to the bit transmission rate. An example of such processing (using Praat [70]) is shown in Figure 10, for a sequence of five 'seek' pulses filled with idle gaps, emitted by the source within a distance of one meter from the receiver. Figure 10,a shows the original waveform. The same waveform after band-pass filtering is seen in Figure 10,b. Finally, intensity values for the filtered waveform are shown in Figure 10,c. Intensity values are relative to the auditory threshold levels. Our values which are around this threshold suggest that the signal can still be decoded despite being barely audible.

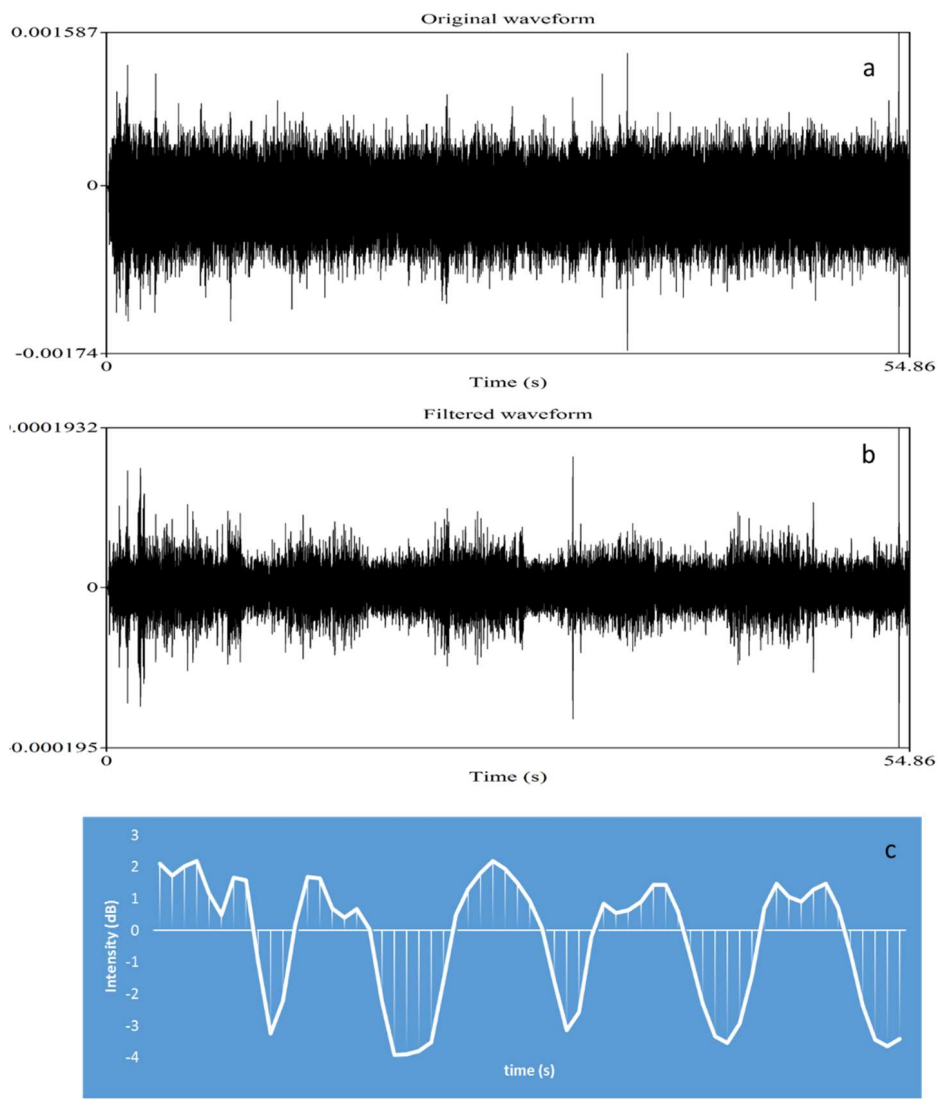

**Figure 10. Waveform of the signal (one meter apart), original (a), filtered (b), and decoded (c)**

### 6.2.2 Receiver Implementation

The acoustic transmissions can be received by a nearby computer with a microphone, a smartphone placed on the desktop, or other types of recording devices. This subsection briefly describes the receiver implementation. Note that audio sampling and on-off keying demodulation are widely used for commutation, and hence are not considered the main contribution of this paper. We refer interested readers to a detailed theoretical explanation and available source-code [71] [72] [73].

We implemented a receiver as an app installed on a Samsung Galaxy S4 (I9500) mobile phone with a standard microphone with a sampling rate of $44.1\ kHz$. The main functionality of the receiver is (1) audio sampling, (2) performing moving windows FFT, (3) preamble detection, and (4) payload demodulation (Algorithm 2).

```
Algorithm 2    Receiver
1:  S = SampleAudio();
2:  FreqWindow = UpdateMovingWindow(S);
3:  if (state == PREAMBLE) then
3:      ChannelProp = DetectPreamble("1010", FreqWindow);
4:      SetState(PAYLOAD);
5:  if (state == PAYLOAD) then
6:      newBit = OOKDemodulate(FreqWindow);
7:      payloadBuffer.addBit(newBit)
8:      if (payloadBuffer.size == 32) then
9:          SetState(PREAMBLE);
10: GOTO 1;
```

The receiver continuously samples the audio signals from the recording device - usually the built-in microphone (line 1). Technically, this is done by utilizing the *AudioRecord* class in the Android framework [74]. We then transfer the signal to the frequency domain using Furrier transform (line 2) [75] . In its PAYLOAD state, the code continuously tries to detect a preamble, by scanning for a sequence of "1010" (a sequence of signal, no signal, signal, no signal). Once payload is detected, the channel properties (e.g., transmission time, noise, etc.) are saved, and the state is set to PAYLOAD (line 4). In a PAYLOAD state, the code demodulates a sequence of 32 bits using the OOK scheme (line 7), then returns to the PREAMBLE state (line 9). Note that error detection and error correction mechanisms, as well as handling of signal loss, are omitted from the algorithm.

## 7. Evaluation

In this section we present the evaluation results based on our experiments and analysis.

**Transmitter**. In our experiments, we used desktop computers installed with the transmitting application as our transmitter. The application can be configured to use 'read,' 'write,' or 'seek' operations, as well as to operate with specified transmission times and predefined sector numbers. During the experiments we checked five different PC desktop workstations with five types of internal HDDs. In addition, we tested *external* HDDs. The list of the computers and HDDs used during the tests is presented in Table 3.

Table 3. Desktop computers and HDD models tested

| # | Type | Model | Chassis |
|---|------|-------|---------|
| **HDD-L** | Internal | Seagate Barracuda 7200.12 ST31000524AS 1TB 7200 RPM 32MB Cache SATA 6.0Gb/s 3.5 Inch | Lenovo |
| **HDD-O** | Internal | WD Blue 1TB Desktop Hard Disk Drive - 7200 RPM SATA 6Gb/s 64MB Cache 3.5 Inch | Optiplex |
| **HDD-A** | Internal | Seagate Barracuda 7200.12 ST3500418AS 500GB 7200 RPM 16MB Cache SATA 3.0Gb/s 3.5 Inch | Antec |
| **HDD-I** | Internal | WD Blue 1TB Desktop Hard Disk Drive - 7200 RPM SATA 6Gb/s 64MB Cache 3.5 Inch | Infinity |
| **HDD-G** | Internal | Seagate Desktop HDD ST1000DM003 1TB 64MB Cache SATA 6.0Gb/s 3.5 Inch | Gigabyte |
| **HDD-EX** | External | WD 500GB drive 2.5" 5400 RPM | - |

During our experiments we did not modify the HDD's automatic acoustic management (AAM) setting, which is usually set on by default. The main reason for this is to keep our covert channel as stealth and quiet as possible in order to evade detection by the user. During all of the experiments the HDDs were firmly installed within computer cases in their usual internal drawers (except for the external HDDs). Before the experiments we validated that the computers' cases were firmly enclosed.

We run the transmitter on desktop computers running the Linux Ubuntu OS, 64-Bit version 14.04.3 kernel 3.13.0. We implemented a version of the receiver as an app for the Android OS. All of our tests were conducted using the Samsung Galaxy S4 smartphone (GT-I9500) installed with stock Android Lollipop (5.0.1). Our testing environment consisted of a computer lab with ordinary background noise, seven workstations, several network switches, and an active air conditioning system.

Figures 11 and 12 show the acoustical waveform generated from HDD-L, as received by a stationary smartphone placed at a distance of one meter and two meters, respectively, from the transmitter. In the two tests we used the 'seek and write' method for the transmission. Using on-off keying modulation, we transmitted a payload of "101010" when $T_0$ = 2 sec and $T_1$ = 1 sec. The received waveform was band-pass filtered between 2050-2100 Hz.

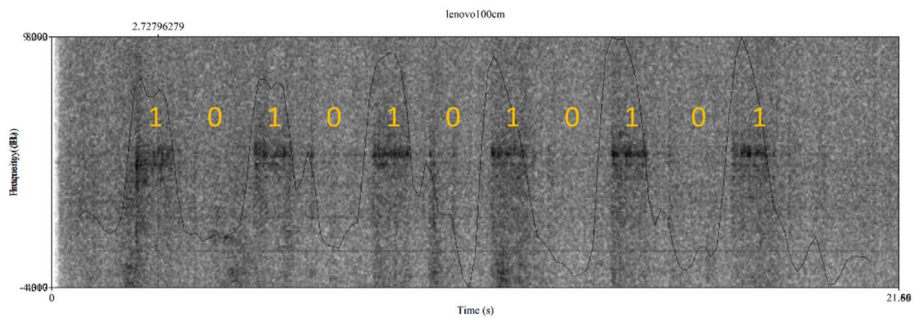

Figure 11. Spectral view of the signal emitted from HDD-L, as received from a distance of one meter

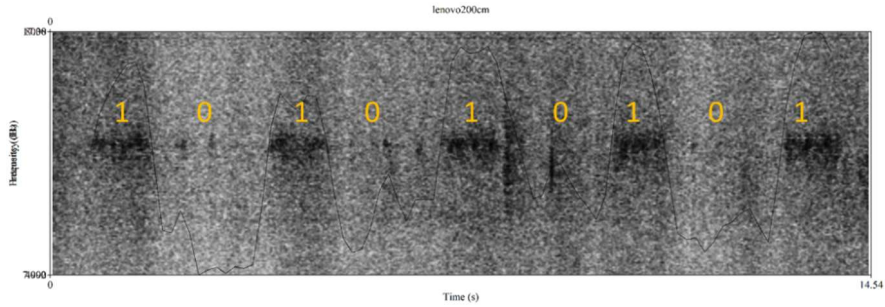

**Figure 12. Spectral view of the signal emitted from HDD-L, as received from a distance of two meters**

Figure 13 shows the acoustical waveforms generated in four tests. HDD-O (Figure 13, a), received by a stationary smartphone placed at a distance of one meter, 'seek and read' method (using dd), and $T_0 = T_1 = 5$ sec. HDD-A (Figure 13, b), received by a stationary smartphone placed at a distance of one meter, 'seek & read' method (using dd), and $T_0 = T_1 = 3$ sec. HDD-I (Figure 13, c), received by a stationary smartphone placed at a distance of one meter, 'seek & read' method (using dd), and $T_0 = T_1 = 3$ sec. HDD-G (Figure 11, d), received by a stationary smartphone placed at a distance of 0.5 meter, 'seek & read' method (using dd), and $T_0 = T_1 = 3$ sec. The received waveform was band-pass filtered between 2050-2100 Hz. In all tests we used on-off keying modulation to transmit a payload of "101010."

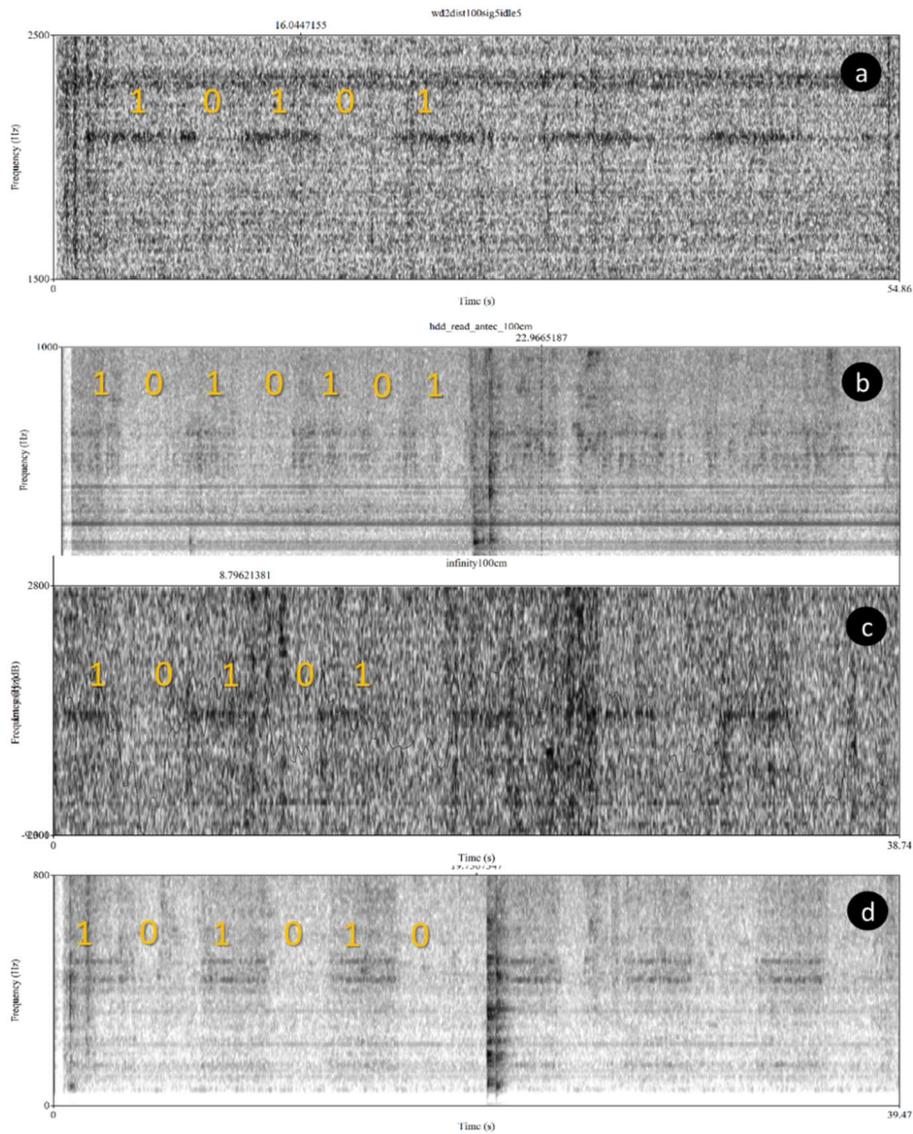

**Figure 13. Spectral view of the signal emitted from four hard drives (HDD-O, HDD-A, HDD-I, and HDD-G)**

Figure 14 shows the acoustical waveform generated from HDD-EX, as received by a stationary smartphone placed at a distance of one meter from the transmitter. In the two tests we use the 'seek and read' method for the transmission. Using on-off keying modulation, we transmitted a payload of "101010" when $T_0 = T_1 = 3$ sec. The received waveform was band-pass filtered between 2050-2100 Hz.

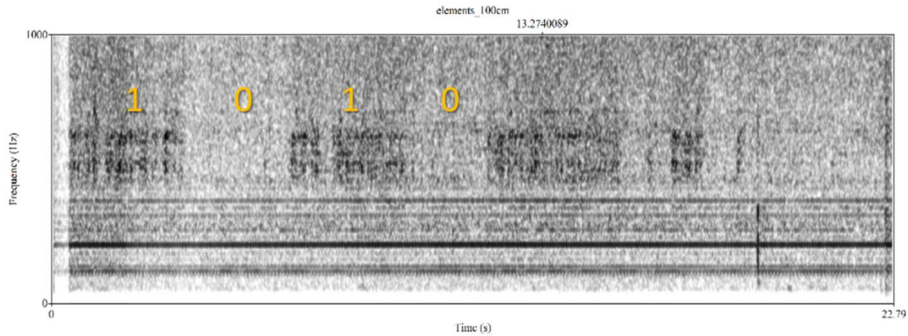

**Figure 14. Spectral view of the signal emitted from HDD-EX at a distance of one meter**

## 7.1 Casual Noise Emission

Since our covert channel is based on HDD activity, casual file operations of other running processes may interfere with the transmissions and interrupt them. Our experiments shows that most applications generate short bursts of noise with only moderate interruptions to the transmission activity. Figure 15 shows the acoustical waveform generated by HDD-I, when the computer was idle, playing video, and performing compilation for a 22 second period

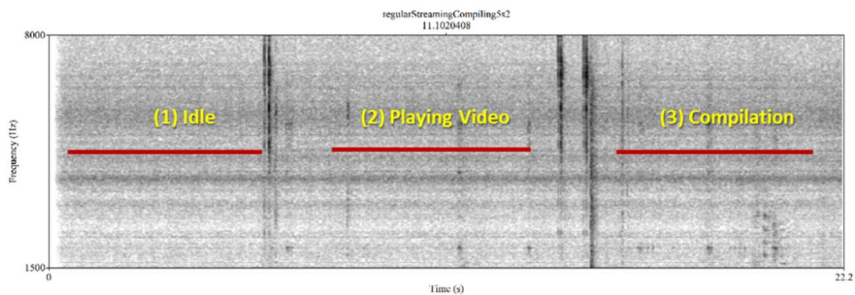

**Figure 15. Spectral view of the signal emitted from HDD-I during different workloads**

As can be seen, the noise generated by casual operations is fleeting in bursts. There are two main reason for this phenomena. First, applications usually read and write files in a sequential manner, sector by sector. This means that most applications rarely seek between different tracks (e.g., files are stored on the same track), which minimizes the acoustic emission from the HDD. Second, the caching mechanisms (in the OS or in the HDD controller) try to reduce the amount of physical access to the hard drive, which minimizes the number of seek operations, and hence the acoustic emissions.

## 7.2 Bit Rate

The bit rate depends on the bit transmission time ($T_0$ and $T_1$) in the on-off keying modulation. Given that $T = T_0 = T_1$, the time it takes to transmit $n$ bits is $n * T$. The values of $T_0$ and $T_1$ may vary, depending on the background noise and the expected distance between the transmitter and the receiver. During our experiments we found that in quiet places, a value of $T_0 = T_1 = 0.3$ seconds allows demodulation with a bit error rate (BER) that is close to 0%. These values of $T_0$ and $T_1$ imply a bit rate of 3 bit/sec = 180 bit/min.

## 8. Countermeasures

Countermeasures to mitigate the DiskFiltration attack can be classified into three categories: hardware based, software based, and procedural based (summarized in Table 5).

**Hardware based countermeasures.** Replacing the HDD drives with SSD can eliminate the threat, since SSDs are not mechanical, hence generating virtually no noises. However, replacement of the hard drive in existing infrastructure may not be always practical due to the high cost [20]. In addition, most PCs, servers, legacy systems, and laptops are still shipped with HDD drives [20]. Acquiring a particularly quiet type of HDD [76] or installing the HDD within special enclosures [77] can also limit the range of emitted noise.

Another type of hardware product includes signal detection and signal jamming systems. Noise detectors [78] aim at monitoring the background noise at specified frequency ranges. However, such noise detectors are usually limited to use in a quiet environment without noise. Jamming the HDD signal by generating static noise in the background is also possible [79] but not particularly applicable in a work environment due to the disturbance it may cause to users.

**Software based countermeasures**. At the software and firmware level, modern HDDs include a feature called automatic acoustic management (AAM) [58] which reduces seek acoustic noise. Ensuring that the AAM settings are at their correct values can limit the range of the emitted signals. As noted, the evaluation in this paper was performed with the default AAM settings, which are configured to their optimal values. Another solution may involve using host intrusion detection systems (HIDS) and host intrusion prevention systems (HIPS) to detect and prevent suspicious 'seek' pattern on HDDs. Such software based countermeasures can be evaded by malware and rootkits at the OS kernel [80] [81]. In addition, distinguishing between legitimate read, write, and seek operations and malicious ones may not be a trivial task.

**Procedural based countermeasures.** Procedural countermeasures involve a physical separation of emanating equipment from potential receivers. This approach is referred to as zone separation by United States and NATO standards [18] [82]. In these standards sensitive computers are kept in restricted areas in which certain equipment is banned. In our case, smartphone and other types of recording devices should not be permitted in close proximity of the computer.

**Table 4. Different types of countermeasures**

| Method | Type |
|---|---|
| **Hardware based** | - Replacing HDDs with SSDs<br>- Acquiring quiet HDDs<br>- Installing special enclosures<br>- Noise detectors<br>- Signal jammers |
| **Software based** | - HIDS/HIPS<br>- Malicious activity detection<br>- Proper configuration of AAM |
| **Procedural** | - Zone separation |

## 9. Conclusion

In this work we present a new type of acoustical out-of-band covert channel code-named DiskFiltration. In this method, an attacker can leak binary data from computers over covert noises emanating from hard disk drives. Unlike most of the existing acoustic covert channels, DiskFiltration can work in computers that are not equipped with speakers or audio hardware. Malicious code installed on the computer can perform intentional seek operations, which cause the HDD head (the actuator) to move between different tracks. The mechanical movements generate acoustic signals which can be used for '0' and '1' modulation. The covert signals can be received by a nearby recording device such as a smartphone, smartwatch, laptop, etc. Despite DiskFiltration's general contribution to the field of covert channels, it is particularly relevant in two adversarial scenarios: (1) in air-gapped networks where there is no network connection between the computer and the Internet, and (2) in computers with heavily monitored (by IDS and IPS systems) Internet connections. In these cases an attacker may resort to out-of-band covert exfiltration channels which are not monitored by existing defense measures. We provided the main technical details regarding the anatomy of modern HDDs and examined the acoustic signals generated by their basic read, write, and seek operations. Based on our observations we designed a rather simple data modulation and demodulation protocol and implemented a prototype of a transmitter (for a computer) and a receiver (for smartphones). We evaluate the covert channel in different types of HDDs and computer chassis, and examine its distance, signal quality, and bandwidth. Finally, we present different types of countermeasures to mitigate this threat. Results shows that DiskFiltration can be used to covertly transfer data to distance of up to two meters (six feet) from the transmitting computer at a bit rate of 180 bits/min (10,800 bits/hour).